\documentclass[]{aa501}
\usepackage{graphics}
\begin{document}

\title{Unusual magnetic fields in the interacting spiral
NGC~3627}

\author{M. Soida\inst{1}
\and M. Urbanik\inst{1}
\and R. Beck\inst{2}
\and R.Wielebinski\inst{2}
\and C. Balkowski\inst{3}}
\institute{Astronomical Observatory, Jagiellonian
University, Krak\'ow,
Poland
 \and Max-Planck-Institut f\"ur Radioastronomie, Bonn,
Germany
 \and Observatoire de Paris, DAEC and UMR 8631, CNRS and
Universit\'e Paris 7, Meudon, France
}
\offprints{M. Soida}
\mail{soida@oa.uj.edu.pl}

\date{Received 3 April 2001; accepted 8 August 2001}

\titlerunning{Unusual magnetic fields in NGC~3627}
\authorrunning{M. Soida et al.}

\abstract{
By observing the interacting galaxy NGC~3627 in radio
polarization we try to answer the question to which degree the
magnetic field follows the galactic gas flows. We obtained total
power and polarized intensity maps at 8.46~GHz and 4.85~GHz using
the VLA in its compact D-configuration. In order to overcome the
zero-spacing problems, the interferometric data were combined
with single-dish measurements obtained with the Effelsberg 100-m
radio telescope. The observed magnetic field structure in
NGC~3627 suggests that two field components are superposed. One
component smoothly fills the interarm space and shows up also in
the outermost disk regions, the other component follows a
symmetric S-shaped structure. In the western disk the latter
component is well aligned with an optical dust lane, following a
bend which is possibly caused by external interactions. However,
in the SE disk the magnetic field crosses a heavy dust lane
segment, apparently being insensitive to strong density-wave
effects. We suggest that the magnetic field is decoupled from the
gas by high turbulent diffusion, in agreement with the large
\ion{H}{i} line width in this region. We discuss in detail the
possible influence of compression effects and non-axisymmetric
gas flows on the general magnetic field asymmetries in NGC~3627.
On the basis of the Faraday rotation distribution we also
suggest the existence of a large ionized halo around this
galaxy.
}

\maketitle
\section{Introduction}

A crucial question in the physics of galactic magnetic
fields is whether the field only passively follows the gas flows
or whether it can develop its own structures driven by small-scale 
turbulent processes (e.g. Donner \& Brandenburg \cite{DB90}, Beck
et al. \cite{rb96}, Moss \cite{mos00}). The first possibility implies that the
magnetic diffusion is very low and the magnetic field remains
perfectly frozen into the gas, following precisely its motions.
In the second case, substantial turbulent diffusion (several
orders of magnitude higher than the Ohmic diffusion) must exist,
allowing large-scale field structures to be built from small-scale 
perturbations caused by turbulent gas motions. Observations
of normal spirals do not provide a clear discrimination between
these two possibilities. Most of the spiral galaxies show regular
magnetic fields with a strong radial component (Beck et al. \cite{rb96})
resisting the differential rotation shear, which indicates a
loose connection of the magnetic field to the gas. In this case
the radial field component would be produced by the dynamo
mechanism (Beck et al. \cite{rb96}), without any need of density wave
flows (see Knapik et al. \cite{jk00}). On the other hand, Neininger \&
Horellou (\cite{NH96}) identified in M51 a magnetic field component
clearly associated with local compression effects (traced by
heavy dust lanes) as expected for a frozen-in magnetic field.
Even more pronounced association of magnetic fields with large-scale 
gas flows was found by Beck et al. (\cite{rb99}) in the strongly
barred galaxy NGC~1097, in which the orientation of the magnetic
field shows rapid changes, as expected for a large-scale bar-induced 
shock. However, the fact that the field lines are bending
some 1~kpc {\it upstream} of the suspected location of the shock
in the bar remains inexplicable in terms of classical MHD models.

In normal, symmetric spirals the dynamo process produces a spiral-like 
magnetic field with a pitch angle similar to that of the
spiral arms, thus almost parallel to the density wave shocks
making their effects upon the field structure very difficult to
observe.  For this reason we performed a two-frequency high-resolution 
study of NGC~3627, an Sb galaxy with peculiar,
asymmetric spiral structure, interacting tidally within the Leo
Triplet group and suspected to have also a perturbed velocity
field (see e.g. Haynes et al. \cite{hay79}). The galaxy is described in
more detail in our earlier low-resolution study of this object
(Soida et al. \cite{ms99}). It shows a bar and two asymmetric spiral
arms with strong density wave effects. While the western arm is
accompanied by weak traces of star formation, the eastern arm
contains a straight actively star-forming segment in its inner
part (see Fig.~\ref{tppi6}). NGC~3627 possesses X-ray properties
of a galaxy with a recent starburst (Dahlem et al. \cite{dal96}), but its
far-infrared emission (more sensitive to the actual content of
young stars) is similar to that of normal spirals (Hughes et al.
\cite{hug91}). Our previous study clearly revealed the existence of a
widespread interarm regular field in addition to possible field
compression effects. Here we discuss in detail the magnetic field
structure in NGC~3627 with a beam size six times smaller than in
our previous work. The use of two frequencies enables to
determine the Faraday rotation distribution from which the global
magnetic field structure may be derived.

\section{Observations and data reduction}

The observations of total power and linearly polarized radio
emission from NGC~3627 were performed in November 1997 at
8.46~GHz and 4.85~GHz using the Very Large Array (VLA) of the
National Radio Astronomy Observatory (NRAO)\footnote{NRAO is a
facility of National Science Foundation operated under
cooperative agreement by Associated Universities, Inc.}. The most
compact D-array configuration was used. The total observation
time (including calibration) was 14~hours at 8.46~GHz and
10~hours at 4.85~GHz.

The intensity scale was calibrated by observing 3C286, adopting
the flux densities of Baars et al. (\cite{baa77}). The position angle of
the linearly polarized emission was calibrated using the same
source with an assumed position angle of 33$\degr$. The phase
calibrator 1117+146 was used to determine the telescope phases
and the instrumental polarization. To check these calibration
procedures 3C138 was observed once each day.

The data reduction has been performed using the AIPS data
reduction package. The edited and calibrated visibility data were
Fourier transformed to obtain the maps of Stokes parameters I, Q
and U at both frequencies. The maps, uniformly weighted (with
ROBUST=0) to obtain the best resolution and sensitivity
compromise, have a synthesized beam with half-power width (HPBW)
of 11$\arcsec$ and 13$\farcs$5 at 8.46~GHz and 4.85~GHz,
respectively. Naturally weighted maps at 8.46~GHz with a
resolution (after a slight convolution) identical to that at
4.85~GHz (13$\farcs$5) were also obtained. The Q and U maps were
combined to get distributions of the linearly polarized emission
(corrected for the positive zero level offset) and of the
position angle of the magnetic vectors (B-vectors).

To increase the sensitivity to extended structures we combined
the data at 8.46~GHz with our earlier observations at 10.55~GHz
made with the 100~m Effelsberg radio telescope (Soida et al.
\cite{ms99}). The brightness values at 10.55~GHz were rescaled to
8.46~GHz assuming a spectral index of 0.64. At 4.85~GHz we
performed new observations with the Effelsberg telescope. We
obtained 11 azimuth-elevation maps of NGC~3627, then combined
into the distributions of Stokes parameters I, U and Q. They were
merged with our VLA maps. At both frequencies we recovered about
10\% of the total power flux and about 5\% polarized flux lost
due to the missing zero-spacing effect.

\begin{figure}[htbp]
\resizebox{\hsize}{!}{\includegraphics{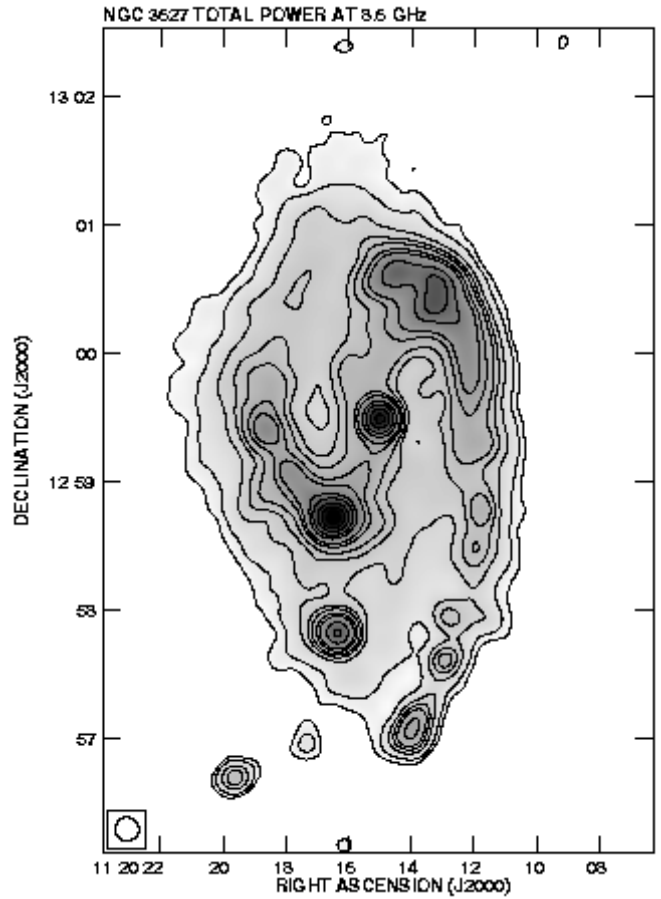}}
\caption{
The combined VLA and Effelsberg total power map of NGC~3627 at
8.46~GHz with uniform weighting providing the highest resolution
of 11$\arcsec$ (contours and greyscale). The contour levels are
(3, 5, 10, 15, 20, 30, 70, 100, 150 200)*27~$\mu$Jy/b.a..
}
\label{tphig}
\end{figure}

Faraday rotation measures in NGC~3627 scarcely exceed
150~rad/m$^2$ (see Fig.~\ref{rm}) which corresponds to a
polarization angle offset of $4\degr$ between 10.55~GHz and
8.46~GHz. The highest RM values coincide with localized peaks of
polarized intensity where the combination of Effelsberg and VLA
data adds little extended emission. Moreover, adding the single
dish data increased the integrated polarized flux only by some
5\%. For these reasons the Faraday rotation between the discussed
frequencies have no effect upon the process of combining VLA and
Effelsberg polarization maps.

\begin{figure}[htbp]
\resizebox{\hsize}{!}{\includegraphics{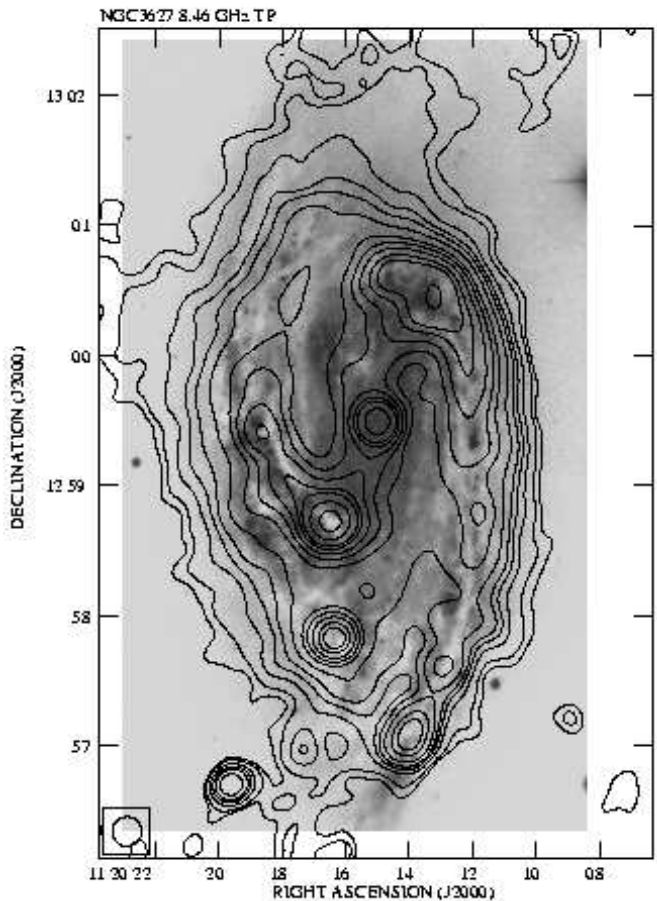}}
\caption{
The combined VLA and Effelsberg total power map of NGC~3627 at
8.46~GHz with a natural weighting (resolution 13$\farcs$5)
superimposed on the optical image from Arp (\cite{arp66}). The contour
levels are (3, 5, 10, 15, 20, 30, 50, 70, 100, 150, 200, 350,
600, 900)*12~$\mu$Jy/b.a.. The map resolution is $13\farcs$5
}
\label{tppi3}
\end{figure}

\section{Results}

\subsection{Total power emission}

\begin{figure}[htbp]
\resizebox{\hsize}{!}{\includegraphics{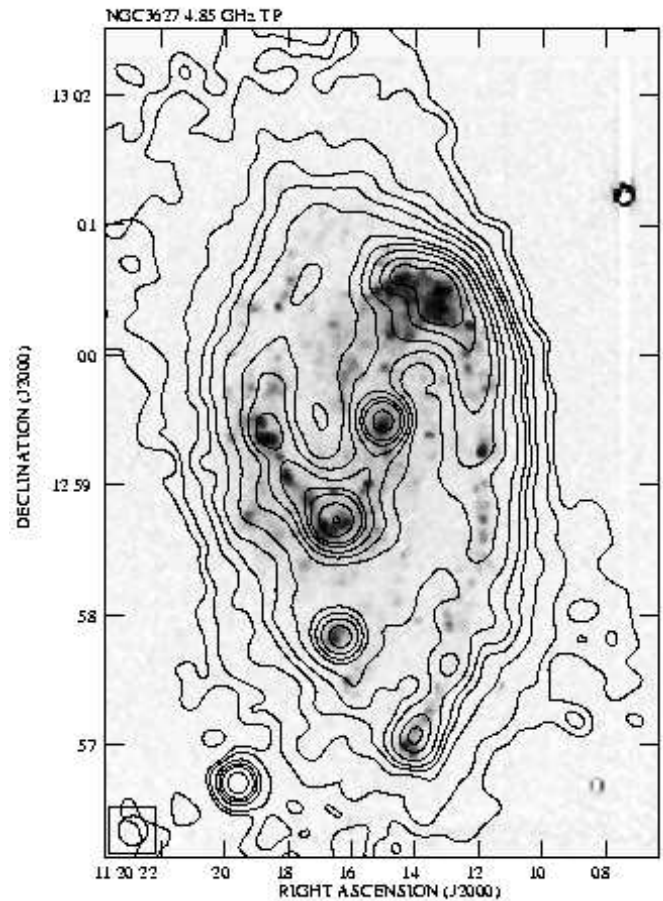}}
\caption{
The combined VLA and Effelsberg total power map of NGC~3627 at
4.85~GHz superimposed upon the H$\alpha$ image from Smith et al.
(\cite{smi94}). The contour levels are: (3, 5, 10, 15, 20, 30, 50, 70,
100, 150, 200, 350, 600, 900)*20~$\mu$Jy/b.a.. The resolution is
13$\farcs$5
}
\label{tppi6}
\end{figure}

The total power map at 8.46~GHz at the resolution of 11\arcsec{}
clearly shows two emission components: an S-shaped ridge
associated with spiral arms and the bar as well as more diffuse
interarm emission (Fig.~\ref{tphig}). In both halves (NW and
SE) of the bar the inner bright ridge is displaced towards the
leading edges with respect to rotation. On the eastern disk side
the total power ridge runs parallel to the heavy dust lane and to
the star-forming, H$\alpha$-bright chain (Figs.~\ref{tppi3} and
\ref{tppi6}), visible also in the distribution of CO and dust
(Reuter at al. \cite{reu96}, Sievers at al. \cite{sie94}). The total power ridge
running along the western arm coincides with both the chain of
\ion{H}{ii} regions (Fig.~\ref{tppi6}) and the dust lane
(Fig.~\ref{tppi3}). Its localized maxima only partly correspond
to individual H$\alpha$ peaks. Bright total power peaks also
coincide with the galaxy centre, the bar ends, with a large
\ion{H}{ii} complex in the southern disk at R.A.$_{2000}=11^h
20^m 16^s $  Dec$_{2000}=+12\degr 57\arcmin 45\arcsec$ and with
another star-forming clump at the southern tip of the western
arm. A very steep brightness gradient occurs at both frequencies
along the NW disk boundary, despite the high sensitivity of the
map at 4.86~GHz, showing well the diffuse emission in the eastern
disk outskirts, including extensions to the north and NE.

\begin{figure}[htbp]
\resizebox{\hsize}{!}{\includegraphics{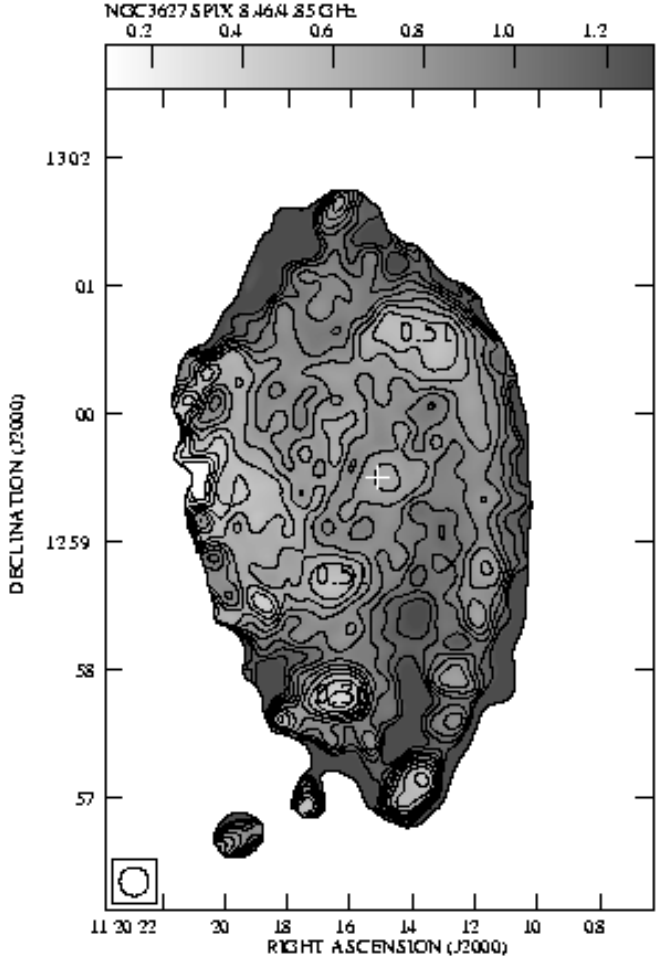}}
\caption{
The distribution of spectral index between 8.46~GHz and 4.85~GHz
in the disk of NGC~3627 at a resolution of 13$\farcs$5. The
contour levels are (1, 2, 3, 4...)*0.1.
Numbers inside the Figure show the values of spectral index in
selected regions of spectral flattening
}
\label{spix}
\end{figure}

The spectral index distribution over the disk of NGC~3627 is
shown in Fig.~\ref{spix}. Apart from some strong variations close
to the disk edges, caused partly by a still remaining slight zero-spacing 
problem, the spectral index $\alpha$ flattens to some 0.5--0.6 
(S $\propto \nu^{-\alpha}$) in star-forming regions at the
bar ends and in the star-forming chain in the eastern disk. A
particular flattening to 0.3--0.4 is observed at the position of
the southern star-forming complex at R.A.$_{2000}=11^{h} 20^{m}
16^s $ Dec$_{2000}=+12\degr 57\arcmin 45\arcsec$ and at the
southern end of the western arm. The spectral index in the inner
disk is about 0.7--0.8 and the same values are found along the
western total power ridge with only slight flattening (to some
0.68) at the positions of \ion{H}{ii} regions. A significant
steepening to about 1.0--1.1 occurs in the interarm region SW of
the centre.

\subsection{Polarization}

\begin{figure}[htbp]
\resizebox{\hsize}{!}{\includegraphics{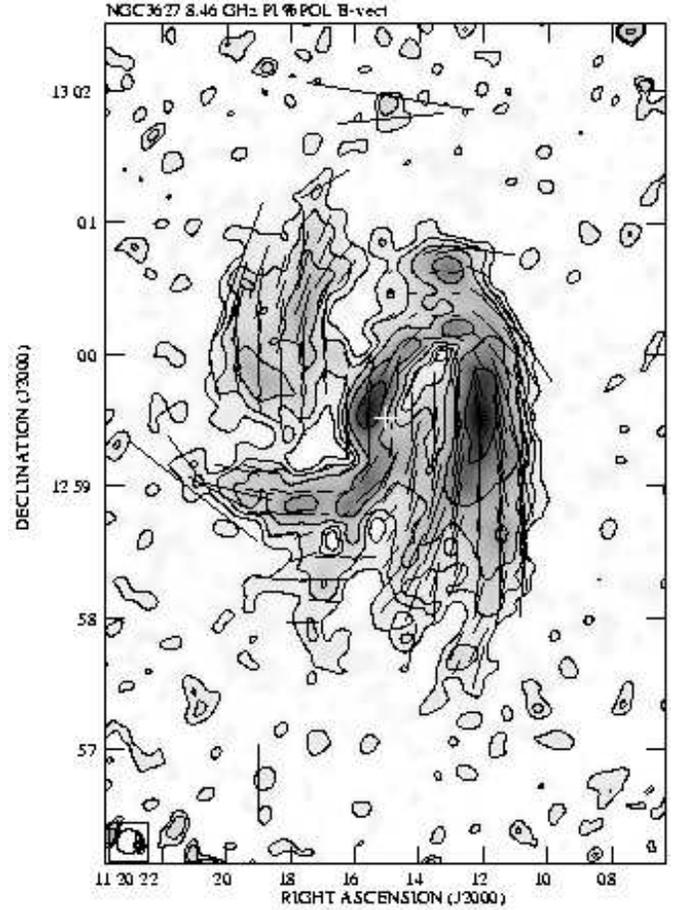}}
\caption{
The contours and greyscale of polarized intensity of NGC~3627 at
8.46~GHz with B-vectors of length proportional to the
polarization degree at the resolution of $11\arcsec$. The contour
levels are (3, 5, 8, 13, 21)*8~$\mu$Jy/b.a.. Apparent B-vectors 
without correction for Faraday rotation are plotted. A
polarization vector equal to $1\arcmin$ length corresponds to a
polarization degree of 60\%. A white cross denotes the optical
disk centre
}
\label{pigrau3}
\end{figure}

The brightest polarized features at 8.46~GHz (Fig.~\ref{pigrau3})
are found in the western disk and northeast of the nucleus. A
very bright polarized peak is located in the middle portion of
the western dust lane. The polarized intensity makes an S-shaped
ridge with B-vectors aligned along it. In the western disk the
polarized ridge crosses the mentioned bright peak of polarized
intensity and extends further southwards along the western dust
lane. In addition, broad polarized features fill the interarm
space in the NE and SW disk (as suggested by Soida et al. \cite{ms99}).
The polarization in the SW interarm region is contaminated by
extensions from the discussed bright polarized peak in the
western spiral arm.

\begin{figure}[htbp]
\resizebox{\hsize}{!}{\includegraphics{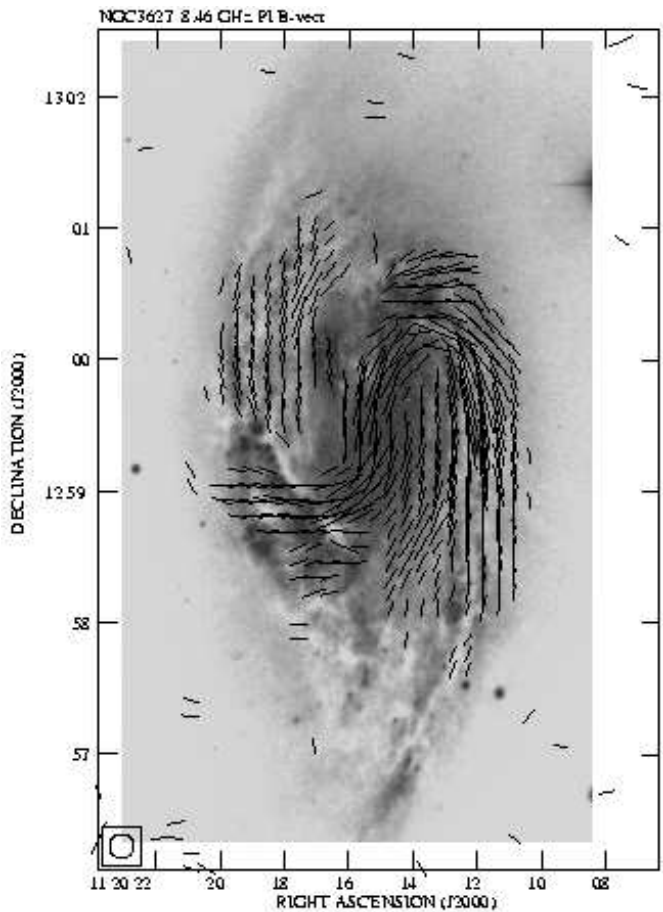}}
\caption{
The B-vectors of length proportional to polarized intensity of
NGC~3627 at 8.46~GHz superimposed onto the optical image by Arp
(\cite{arp66}). A vector length of 10\arcsec corresponds to a polarized
intensity of 50~$\mu$Jy/b.a. The orientations of vectors have
been corrected for Faraday rotation effects. The map resolution
is $11\arcsec$
}
\label{vec3}
\end{figure}

In the western disk the B-vectors at 8.46~GHz run parallel to the
dust lane (Fig.~\ref{vec3}), but in the interarm region they make
a smooth transition to the eastern arm. However, in the vicinity
of R.A.$_{2000}=11^{h} 20^{m} 13\fs 2 $ Dec$_{2000}=+13\degr
00\arcmin $ the B-vectors in the S-shaped feature (above this
position) and in the interarm region (below this position) have
orientations differing by about $90\degr$. This occurs some
$11\arcsec$ (a full beam size) south of the dust lane and cannot
result from pure resolution effects. It also gives rise to beam
depolarization causing the unpolarized ``hole'' at the indicated
position. The difference in position angles between the interarm
region and the S-shaped feature becomes even more impressive
after deprojection to the face-on view (Fig.~\ref{face}). In
the eastern disk, the S-shaped ridge and the B-vector
orientations {\it run across} the optical arm and a heavy dust
lane at a high angle. This is rarely observed in normal
galaxies until now. The B-vectors in this ``magnetic arm'' also
meet at right angles with the interarm vectors, which corresponds
to a depolarized region observed at low resolution by Soida et
al. (\cite{ms99}).

\begin{figure}[htbp]
\resizebox{\hsize}{!}{\includegraphics{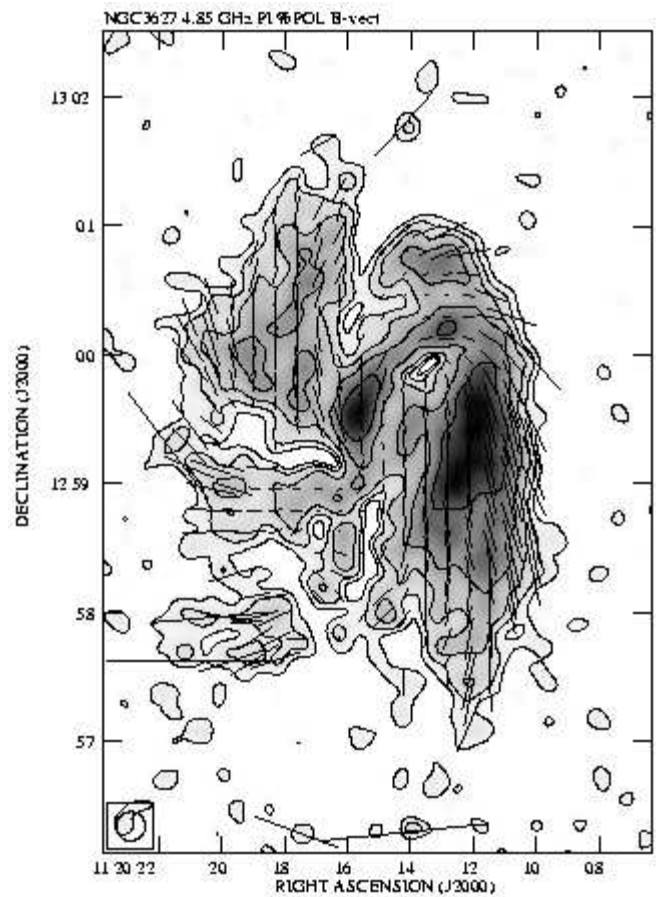}}
\caption{
The contours and greyplot of the polarized emission from NGC~3627
at 4.86~GHz (resolution of $13\farcs$5) with superimposed vectors
of polarization degree. The combined VLA and Effelsberg map is
used. The contour levels are (3, 5, 8, 13, 21)*10 ~$\mu$Jy/b.a..
A polarization vector equal to $1\arcmin$ length corresponds to
the polarization degree of 60\%
}
\label{pi6opt}
\end{figure}

The more sensitive polarization map at 4.85~GHz
(Fig.~\ref{pi6opt}) reveals a regular spiral structure in the
outer disk of NGC~3627. This is visible along both the western
(north and south of R.A.$_{2000}=11^{h} 20^{m} 11^s $
Dec$_{2000}=+12\degr 59\arcmin $) and eastern (around
R.A.$_{2000}=11^{h} 20^{m} 20^s $ Dec$_{2000}=+13\degr 00\arcmin
$) disk boundary where the B-vectors form a regular spiral
structure considerably inclined with respect to adjacent
dust lanes. The polarized region in the SE disk
(R.A.$_{2000}=11^{h} 20^{m} 20\arcsec$ Dec$_{2000}=+12\degr
57\arcmin 45\arcsec$) with magnetic vectors bent outwards was
found to extend beyond the optical image (compare Figs.~\ref{pi6opt}
and~\ref{vec3}). It is particularly well visible
when single-dish Effelsberg data are added.

\subsection{Faraday rotation}

\begin{figure}[htbp]
\resizebox{\hsize}{!}{\includegraphics{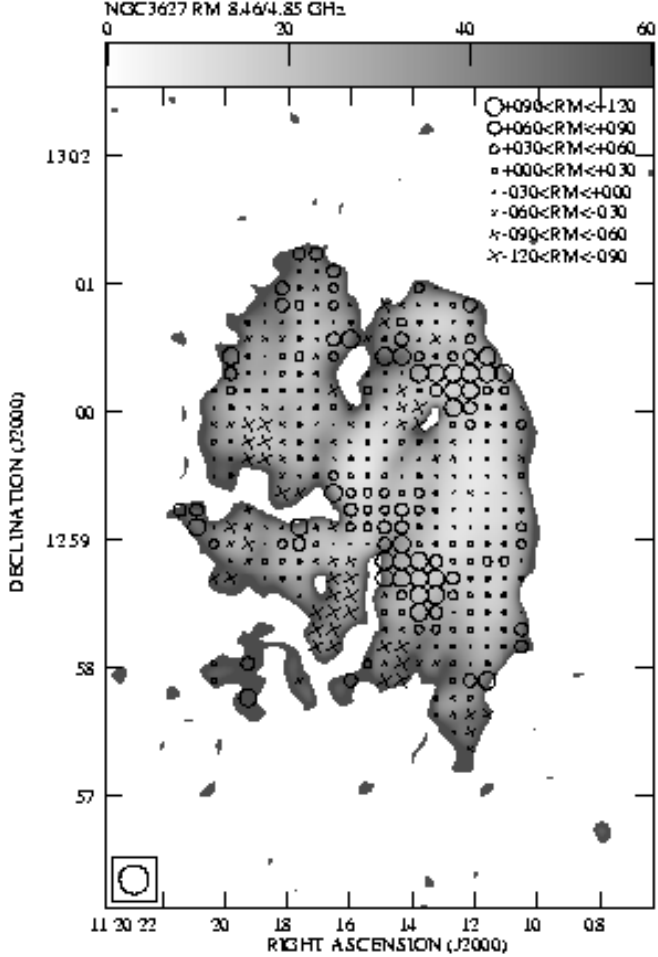}}
\caption{
The distribution of Faraday rotation between 8.46~GHz and
4.85~GHz in the disk of NGC~3627 presented in the form of symbols
with sizes corresponding to particular RM ranges (see the Figure
legend). At both frequencies maps with a resolution of
13$\farcs$5 were used. The greyscale plot shows the RM errors due
to noise
}
\label{rm}
\end{figure}

The map of Faraday rotation measures (RM) between 8.46~GHz and
4.85~GHz is shown in Fig.~\ref{rm}. A foreground Faraday rotation
correction of 50~rad/m$^2$ has been applied as a best guess
between values for that region given by Simard-Normandin \&
Kronberg (\cite{SNK80}) and the mean RM for the whole disk of NGC~3627.
Regions of large positive RM ($> 150$~rad/m$^2$) were found close
to the star-forming region in the northern bar portion and in the
interarm region in the SW disk. Large negative RMs (absolute
values in excess of 150~rad/m$^2$) extend between the southern
bar end and the \ion{H}{ii} complex at R.A.$_{2000}=11^{h} 20^{m}
16^s $ Dec$_{2000} = +12\degr 57\arcmin 45\arcsec$. They also are
present in the SW disk where a sudden change of RM signs takes
place. Apart from that the Faraday rotation measures are
generally small, scarcely exceeding $\pm 50$~rad/m$^2$. They are
changing sign along spiral arms over regions several kpc in size,
not providing any clear picture of a single global magnetic field
symmetry. The ``magnetic arm'' crossing the optical dust lane in
the eastern disk also shows the reversal of RM sign.

\section{Discussion}

\subsection{ Thermal and nonthermal emission}

The radio spectrum of a given region of the galactic disk
results from a combination of diffuse emission from nonthermal
electrons filling the whole disk, nonthermal emission from young
supernova remnants (SNR), and free-free emission from the ionized
gas, both usually concentrated along the spiral arms. The
spectrum of the diffuse component is determined by the balance
between the propagation effects and synchrotron losses (Lerche \&
Schlickeiser \cite{LS81}), it is usually assumed to vary little across
the disk (van der Kruit \cite{kru77}, Hummel \& Gr\"ave \cite{HG90}, Ehle \&
Beck \cite{EB93}). The bright total power peaks correspond well to
maxima visible in the H$\alpha$ map (Fig.~\ref{tppi6}).

\begin{figure}[htbp]
\resizebox{\hsize}{!}{\includegraphics{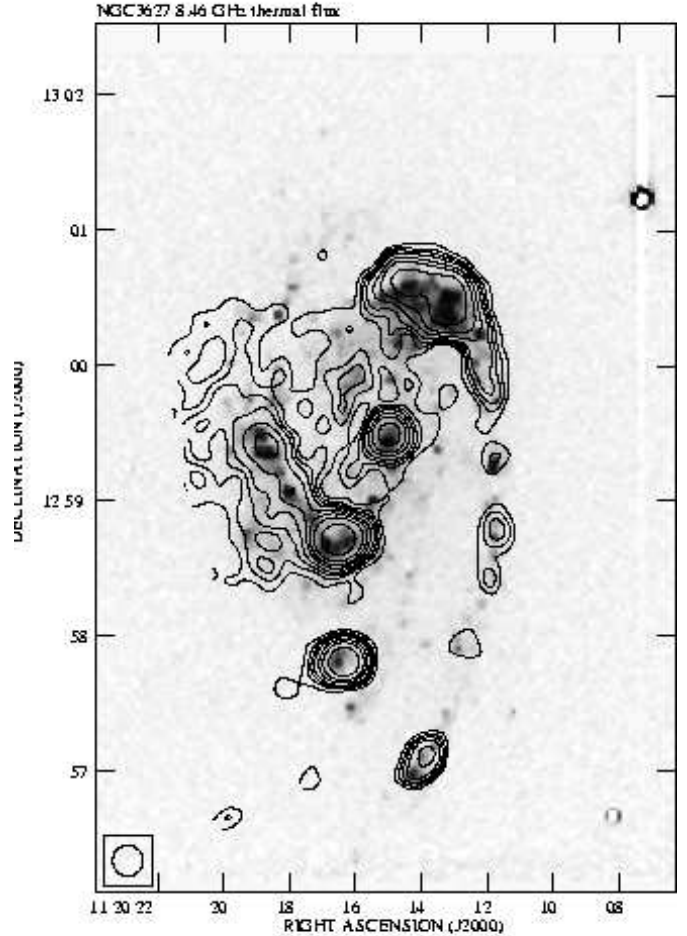}}
\caption{
The overlay of the map of thermal flux at 8.46~GHz obtained under
the assumption of a constant nonthermal spectral index of 0.9,
onto the H$\alpha$ image from Smith et al. (\cite{smi94}). The contour
levels are (2, 5, 8, 12, 20, 30, 50, 100)*20~$\mu$Jy/b.a.
}
\label{therm}
\end{figure}

According to the above we interpret the spectral flattening in
the total power peaks as being due to an increased fraction of
thermal emission mixed with contributions from young SNR's, which
is usually several times smaller than that from the free-free
emission (Urbanik \cite{urb87}). Assuming a mean nonthermal spectral
index of 0.9 (the mean value in the NE disk, away from star-forming 
complexes) and neglecting the SNR contribution, the
thermal fraction f$_{th}$ at 8.46~GHz is about 40\% in the bar
ends and $\simeq$ 60\% in the southern star-forming complex
mentioned in Sect. 3.1. There is a very good correspondence
between the thermal brightness derived from the spectrum and the
distribution of H$\alpha$ emission (Fig.~\ref{therm}).

The knowledge of nonthermal intensity enables us to estimate
the magnetic field strength in specific regions of NGC~3627. We
assumed a nonthermal spectral index of 0.9, a face-on nonthermal
disk thickness of 2~kpc, a lower electron energy cutoff at
300~MeV and a proton-to-electron ratio of 100. The errors of the
field strength include the uncertainty in the last three
quantities of a factor of 2. We computed pressure-balance
magnetic field strengths constituting the minimum field strength
below which the mixture of cosmic rays and magnetic field becomes
unstable. For the mean magnetic field inside the contour level of
$90\mu$Jy/b.a. (1\% of the maximum brightness) we obtained $11\pm
2~\mu$G for the total field. From the polarized intensity in this
area we found $4\pm 1~\mu$G for the regular magnetic field. For
the interarm regions we obtained: $10\pm 3~\mu$G for the NE and $
10\pm 2~\mu$G for the SW interarm regions. We also computed the
mean pressure-balance magnetic field strengths averaged along the
spiral arms. Along the western arm we obtained $11\pm 3~\mu$G and
along the eastern arm $12\pm 3~\mu$G, again with no significant
difference. For the regular fields we obtained $5\pm 1~\mu$G for
the western arm and $4\pm 1~\mu$G for the eastern one thus a
marginally stronger {\it pressure balance} regular field in the
western arm. We note, however, that these values represent the
minimum field strength ensuring its stable configuration for a
given synchrotron brightness. While this is probably valid on
average in the disk, the magnetic field may be stronger in
specific regions than expected from pressure balance, without
causing a stability problem.

We verified our estimates by predicting a Faraday rotation
in the region around  R.A.$_{2000}=11^{h} 20^{m} 12\fs 6 $
Dec$_{2000}=+13\degr 00\arcmin 16\arcsec$, where the regular
field runs along the minor axis, mostly parallel to the line of
sight. The measured RM in this area is measured $130\pm
16$~rad/m$^2$ while its thermal brightness is 0.48~mJy/b.a.. To
reproduce consistently these values we assumed that 20\% of
thermal emission originates in the diffuse gas responsible for
Faraday effects (Walterbos \& Braun \cite{WB94}) which has a face-on
full thickness of 1400~pc (Fletcher et al. in prep.). We obtained
a filling factor of the diffuse gas of 0.025, thus similar to the
value suggested for the Milky Way by Fletcher et al. (in prep.),
implying a peak electron density n$_0\approx 1.1$~cm$^{-3}$
corresponding to a mean electron density along the line of sight
of 0.03~cm$^{-3}$. This is comparable to the value usually
adopted for our Galaxy.

\subsection{Compression effects of the gas and magnetic
field }

The western total power ridge has a typical nonthermal
spectrum. Under the above assumption the mean thermal fraction at
8.46~GHz and its r.m.s. variation along the ridge is $9\pm 5\%$.
It also has a counterpart in the polarized emission and we can
largely attribute it to effects of magnetic field enhancement in
the dust lane. The same is true for the total power bridge
crossing the centre of NGC~3627, whose distortion toward the
leading bar edges is suggestive for compression effects on the
front side of the bar (e.g. Patsis \& Athanassoula \cite{PA00}). We also
see some polarized emission extending beyond the northern bar
end, to the side opposite of the arm, resembling the ``T-regions'' 
discussed by Patsis \& Athanassoula (\cite{PA00}).

\begin{figure*}[htbp]
\resizebox{12cm}{!}{\includegraphics{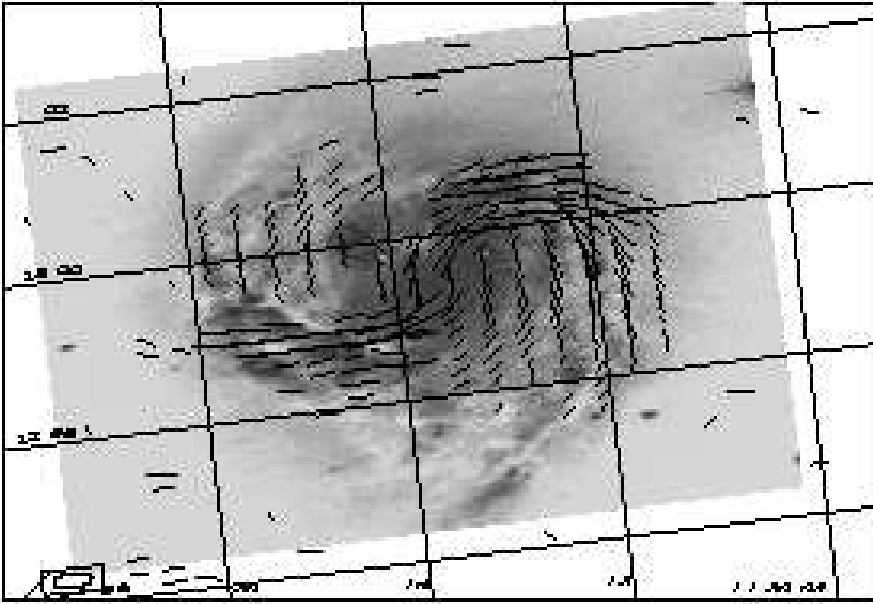}}
\parbox[b]{55mm}{
\caption{
The orientation of B-vectors of polarized intensity at 8.46~GHz
compared to the optical image of Arp (\cite{arp66}), rectified to the
face-on position (resolution of $11\arcsec$)
}
\label{face}}
\end{figure*}

The western disk boundary shows some signs of gas and magnetic
field compression by external influences. We note a steep total
power gradient and a bright polarized peak in the western disk.
There is another (weaker) polarization peak just adjacent to
the nucleus (Fig.~\ref{pigrau3}), it is probably associated with
compression by shocks occurring in the central part of the bar.
This has been observed in barred spirals (Beck et al. \cite{rb99}) while
a very bright, elongated polarized area on one side of the galaxy
is rather uncommon. It is coincident with a ``pushed inwards''
segment of the dust lane at an azimuth angle of 280\degr,
$\ln(r)\approx 0.4$ (Fig.~\ref{mu}), followed by a similar
deviation of magnetic pitch angles, it is also accompanied by a
steep \ion{H}{i} density gradient along the western disk side
discussed by Haynes et al. (\cite{hay79}).

We note that the SW disk region (west and south of the
nucleus) has a mean {\it nonthermal} brightness by about 10\%
higher than the NE disk while its mean star formation level (as
measured by a thermal flux) is about 3 times lower. The western
disk portion may thus have the magnetic field relative to its
star formation level considerably stronger than the eastern one.
This is suggestive for some external field enhancement (e.g. by
compression).

All these phenomena occurring together may constitute a signature
of compression effects, possibly caused by the encounter with the
gas tidally stripped during interactions with NGC~3628 (see e.g.
numerical models by Rots (\cite{rot78})). However, a simple picture of
gas and magnetic field compression does not explain large
magnetic pitch angles observed westwards of the western dust lane
(upstream of a possible flow, north and south of
R.A.$_{2000}=10^{h} 22^{m} 10^s $ Dec$_{2000}=+12\degr
59\arcmin$, see Fig.~\ref{pi6opt}). The magnetic pitch angles
are considerably greater than those of the adjacent dust lane. A
better explanation may involve the magnetic field connected to
dense gas in the western dust lane detected in CO by Reuter et
al. (\cite{reu96}). In the warmer medium outside the lane the pitch angle
could be determined by the dynamo process while the dust lane
formation (distorted by the gas infall) locally aligns the
magnetic field with the lane. A detailed test will be possible
once the kinematical properties of various gas phases with a few
arcsec resolution are known.

\subsection{Two magnetic field components in NGC~3627?}

\begin{figure*}[htbp]
\resizebox{12cm}{!}{\includegraphics{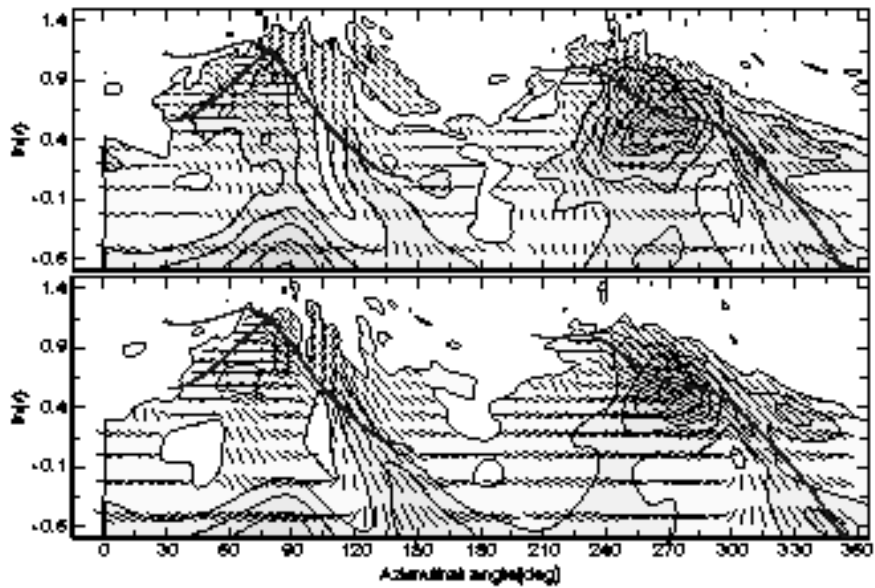}}
\parbox[b]{55mm}{
\caption{
The distribution of polarized intensity and B-vector orientations
at 4.85~GHz (upper panel) and at 8.46~GHz (lower panel) in the
frame of azimuthal angle in the disk and $\ln(r)$. The azimuthal
angle runs anticlockwise from the southern major axis end. The
maps at the resolution of $13\farcs$5 were used. Thick lines
denote the optically visible dust lanes
}
\label{mu}}
\end{figure*}

The polarized component associated with the bar and spiral arm
forms an S-shaped polarized ridge which is quite symmetric in the
inner part (Fig.~\ref{face} and~\ref{mu}). At larger radii
its western part follows the ``pushed-in'' western arm and its
sharply bending outer segment. The eastern portion preserves a
constant large pitch angle, remaining insensitive to any effects
of the gas compression. The latter property is uncommon among
normal spirals in which the magnetic field is usually parallel to
the dust lanes. Another known example of a magnetic arm crossing
the optical one is NGC~2997 (Han et al. \cite{han99}). Apparently, the
gas flow in the bar generates its own magnetic field component,
which reacts in a different way to the compression effects in the
NW and in SE spiral arm segments. Zhang et al. (\cite{zha93}) observing
the \ion{H}{i} motions at the resolution of $19\arcsec$
noted a kinematical difference between these regions. While the
region of the northern bar end and the NE arm segment shows a
rather regular velocity field, strong gas motion perturbations
occur in the SE disk which are also visible in the CO map by
Reuter et al. \cite{reu96}. We note the counter-rotating plume found
there by Haynes et al. (\cite{hay79}) east of R.A.$_{2000}=11^{h}
20^{m} 20^s $ Dec$_{2000}=+12\degr 59\arcmin$. The \ion{H}{i}
line observed by Zhang et al. (\cite{zha93}) is considerably broader
there as well. We may speculate that an increased turbulent
activity (due e.g. to violent, interaction-induced shearing gas
motions) gives rise to much larger turbulent diffusion and much
weaker coupling of magnetic field to the gas than in the more
quiet NW arm segment. Such a ``decoupling'' has also been
obtained by Elstner et al. (\cite{del00}) in the MHD model with an
increased turbulent diffusion in spiral arms. This effect may
also occur in case of cold, dense gas. Even in case of a low
ionization level the magnetic field may still be tied to the
molecular matter due to efficient collisions between ions and
neutral atoms, still yielding a low diffusion coefficient. In the
SE region of NGC~3627 an increased turbulent diffusion may
liberate the magnetic field from the cold gas which forms the
prominent dust lane segment. An alternative possibility that the
magnetic field remains tied to the gas but the peculiar SE part
of the S-shaped structure is strongly bent away from the plane of
NGC~3627, running high above the dust lane has to be considered,
too.

There are arguments in support of both possibilities.
Assuming that the peculiar arm is in the plane would imply a high
Faraday rotation, as the magnetic field would be mostly parallel
to the line of sight while the underlying star-forming region
supplies enough ionized gas. The problem would disappear if the
magnetic arm were bent strongly out of plane making the regular
field almost perpendicular to the line of sight. One of most
recent determinations of the distance to NGC~3627 is 11.9~kpc
(Ryan and Visvanathan \cite{RV89}). If the polarized arm were inclined
to the plane only by $30\degr$, its eastern part would deviate by
more than 2.5 kpc from the disk plane. Additionally the gas
density (hence Faraday rotation) would be small there. On the
other hand, its polarized synchrotron brightness is the same as
that of its NW counterpart (Fig.~\ref{pigrau3}) which raises some
problems with supplying cosmic ray electrons to such a height
above the disk. In NGC 5775, showing cosmic ray streaming
along regular fields highly inclined to the plane (T\"ullmann et
al. \cite{tul00}) the polarized brightness at 4.86~GHz fades by a factor
of 2.5--3 over the distance of 1.5 kpc from the plane. In NGC
3627 it decreases only by a factor of 0.7 over the length of the
whole polarized arm, though we still cannot reject some ultra-fast 
cosmic-ray electron propagation along this feature. We note
also that, if we assume the magnetic field strongly coupled to
the gas, we need also to explain a complete lack of regular
magnetic fields in a very strong SE dust lane segment. Projected
to the sky plane at almost a right angle to the polarized arm it
would strongly depolarize its middle portion, which is not the
case (Fig.~\ref{pigrau3}). Finally, we note that out to the
radius of $1\farcm 5$ the NW and SE parts of the S-shaped feature
look very symmetric, both in the distribution of polarized
intensity and in the orientation of polarization B-vectors
(Fig.~\ref{mu}), which would need a very special geometry of an
asymmetric structure, three-dimensionally bent out of the
plane. A detailed discussion of the nature of this polarized arm
requires data on gas kinematics of this region with arcsecond
resolution.

The interarm magnetic field SW and NE of the centre (around
R.A.$_{2000}=11^{h} 20^{m} 14\fs 0 $ Dec$_{2000}=+12\degr
59\arcmin$ and R.A.$_{2000}=11^{h} 20^{m} 18\fs 0$
Dec$_{2000}=+13\degr 00\arcmin$) apparently behaves in a way
different from that associated with spiral arms. In the
inner disk $(\ln(r)<0.4)$ the interarm field has an average pitch
angle of $27\pm 2\degr$ in the eastern disk  and $16 \pm 2\degr$
in the western interarm region. This is smaller than the pitch
angle of $\approx 50\degr$--$60\degr$ in the S-shaped component
within the same galactocentric distance (Fig.~\ref{face} and
\ref{mu}). A strong difference in magnetic pitch angles between
the bar/spiral and interarm fields is visible both in the NW and
the SE disk at azimuthal angles of about $100\degr$ and
300\degr. In the SE disk the region along which the B-vectors
differ by $90\degr$ shows no association with the heavy dust
lane. In the NW region the turning of B-vectors occurs some 1~kpc
upstream of the lane. Some departures  from the general rotation
of the CO-emitting gas were found  close to the eastern jump in
the magnetic field direction (Reuter at al. \cite{reu96}). However,
because of limited resolution these residuals may be due to
generally complex gas kinematics in this region (Zhang et al.
\cite{zha93}). No velocity perturbations are seen close to the NW
magnetic field jump. There is no evidence for associations
of sudden turns with any shock-like phenomena which would suggest
that the S-shaped feature and the interarm field constitute the
same magnetic field strongly distorted by density wave flows
(e.g. Otmianowska-Mazur \& Chiba (\cite{OMC95}). We may speculate that
rapid turns in the field direction may constitute intrinsic
magnetic features at the interface of two physically distinct
magnetic field components. One of them is associated with the bar
and spiral arms while the second component is generated by the
turbulent dynamo process acting in the interarm space,
unperturbed by the density waves. Both SW and NE interarm regions
contain some weak dust filaments too. However, in large regions
around R.A.$_{2000} = 11^{h} 20^{m} 14\fs 0$ Dec$_{2000} =
+12\degr 58\arcmin 30\arcsec$ and R.A.$_{2000} = 11^{h} 20^{m}
17\fs 0$ Dec$_{2000} = +13\degr 00\arcmin 45\arcsec$ the B-vectors 
deviate by more than $30\degr$ from underlying dust
structures. It is thus unlikely that the magnetic pitch angle in
the interarm region is determined by local compression effects
traced by the dust filaments.

We also note that the interarm magnetic pitch angles are
somewhat greater in the eastern, uncompressed region (possibly
containing a pulled-out gas, Haynes at al. \cite{hay79}). This bears
resemblance to some wind-swept galaxies like NGC~2276 (Hummel \&
Beck \cite{HB95}) or NGC~4254 (Soida et al. \cite{ms96}). Such a picture is
suggestive for some global non-circular component of the gas
motion in NGC~3627 oriented generally from the NW-W-SW directions
towards the east, as indeed expected from the kinematic picture
discussed by Haynes et al. (\cite{hay79}).

We suggest the following tentative picture of magnetic
fields in NGC~3627:

\begin{itemize}
  \item[-] NGC~3627 contains two magnetic field components:
  \begin{itemize}
  
  \item[a. ] a smoothly distributed component produced by the
    classical dynamo working in the unperturbed way in the
    interarm region. It results in small pitch angles in the
    inner disk and larger pitch angle values in the outer regions
    (less shear by differential rotation),
  
  \item[b. ] a two-arm magnetic wave generated by the interaction
    of the dynamo with bar flows (resonances with the bar, e.g.
    Chiba \cite{chi93}, Moss \cite{mos98}).
  
  \end{itemize}
  
  Both components have different pitch angles with a sharp
    transition at their interface.

\item[-] External interactions cause some asymmetries in the
  spiral field component. In the more quiet cold gas in the
  western arm the field lines locally follow the dust lane. In
  the more turbulent eastern arm the magnetic field preserves its
  own pitch angle independently of the gaseous arm because of
  larger turbulent diffusion.

\item[-] The interaction-induced eastward gas flow causes also
  some asymmetry in the interarm magnetic pitch angles.
\end{itemize}

Unfortunately, little can be said about the global magnetic field
modes. The RM data do not show any specific pattern which could
be identified either with axisymmetric (ASS) or bisymmetric (BSS)
magnetic fields alone. An analysis of magnetic field modes in a
number of concentric rings $15\arcsec$ wide, involving both
Faraday rotation and observed magnetic pitch angles, was kindly
made for us by A. Fletcher (see Fletcher et al. \cite{af00}). No
fit was possible for single modes: axisymmetric (m=0), or
bisymmetric (m=1). Attempts to involve the mixtures of m=0 and 1
or m=0 and 2 gave regular magnetic field strengths and intrinsic
pitch angles differing from ring to ring by a factor of several.
The most reasonable fit is a mixture of m=0, 1 and 2 field modes
in mutual proportions varying from one ring to another. This
resembles situations modeled by Urbanik et al. (\cite{urb97}) when large-scale 
magnetic fields in the halo (invisible in emission because
of lack of CR electrons at large heights) yield patterns in
Faraday rotation very different from what one could expect from
assumed simple field structures. This also has been suggested to
happen in NGC~6946 (Beck \cite{rb91}). We suspect that the complex
Faraday rotation pattern in NGC~3627 may originate in extended
hot, low-density ionized magnetized halo of the galaxy and may
constitute an indirect evidence for large-scale magnetic field
unseen in synchrotron emission.

\section{Summary and conclusions}

We observed the Leo Triplet spiral NGC~3627 with the VLA D-array
at 8.46~GHz and 4.85~GHz. The data were combined with the
Effelsberg single-dish observations at 10.55~GHz and at 4.85~GHz
to increase the sensitivity to extended structures. The following
results were obtained:

\begin{itemize}

\item[-] The total power emission shows good correspondence to
the distribution of star formation with increased thermal
emission in disk regions with strong star formation.

\item[-] The western disk boundary of NGC~3627 shows several
phenomena which are suggestive for the gas and magnetic field
compression by encounters with tidally stripped gas.

\item[-] In the outer disk, west of the western dust lane the
magnetic field seems to form a spiral-like structure with a
magnetic pitch angle greater than that of the nearby optical
structures. We suspect that magnetic field is connected to the
western dust lane distorted by interactions in its middle
portion, rather than being generally compressed along the western
disk edge.

\item[-] We suggest that two distinct magnetic field
components are present in NGC~3627: one component related
to spiral arms and an interarm component. These components
locally show dramatically discrepant magnetic pitch angles. The
rapid changes of the magnetic field orientation are not clearly
associated with optical features which could be attributed to
density wave shocks.

\item[-] The postulated arm component of the magnetic field
apparently undergoes compression along the western disk side but
crosses the eastern arm at a large angle. We suspect an increased
turbulent diffusion in this region.

\item[-] The interarm magnetic field has also larger pitch angles
in the eastern than in the western disk region, suggestive for
eastward non-axisymmetric global gas motions.

\item[-] The global magnetic field does not show a single ASS or
BSS symmetry. A mixture of modes is probable. The existence of a
large magnetized halo of low density hot gas around NGC~3627 is
also suggested.

\end{itemize}

The understanding of our present observations needs a good
knowledge of the kinematics of NGC~3627 with high resolution. The
existing data are insufficient to answer the question whether the
rapid changes in magnetic field orientation are associated with
some features in the gas motions. High-resolution information on
the kinematics of the region where the magnetic field crosses the
optical arm is highly desirable. These problems will be the
subject of a separate, extensive high-resolution study of the
kinematics of various gas phases in NGC~3627 (in preparation).

\begin{acknowledgements}
We are grateful to numerous colleagues from the Max-Planck-Institut 
f\"ur Radioastronomie (MPIfR) in Bonn for their valuable
discussions during this work. We wish to express our thanks to Dr
Andrew Fletcher from the University of Newcastle for his
calculations of magnetic field modes and to Dr Beverly Smith from
the Lovell Observatory, for providing us with an excellent
H$\alpha$ map of NGC~3627. We are indebted to Dr Elly M.
Berkhuijsen from MPIfR for her help in improving the manuscript
of this paper. One of us (MU) is indebted to the University Paris
7 for obtaining a visiting position. We are also grateful to
colleagues from the Astronomical Observatory of the Jagiellonian
University in Krak\'ow for their comments. This work was
supported by a grant from the Polish Research Committee (KBN),
grant no. 4264/P03/99/17. Large parts of computations were made
with the Convex-SPP machine at the Academic Computer Centre
``Cyfronet'' in Krak\'ow (grant no. KBN/SPP/UJ/011/1996).
\end{acknowledgements}

\end{document}